\documentclass[reprint]{revtex4-1}
\usepackage{bbm}

\usepackage{amsmath}
\usepackage{braket}
\usepackage[utf8]{inputenc}
\usepackage[version=4]{mhchem}
\usepackage[thinc]{esdiff}
\usepackage{xargs}                      
\DeclareUnicodeCharacter{2009}{ }
\usepackage{graphicx}
\usepackage{subcaption}

\newcommand{\ah}{\hat{a}}
\newcommand{\ahd}{\hat{a}^\dagger}
\newcommand{\bh}{\hat{b}}
\newcommand{\bhd}{\hat{b}^\dagger}
\newcommand{\HH}[1]{\hat{H}_\text{#1}}
\newcommand{\at}{\tilde{\alpha}}
\newcommand{\bt}{\tilde{\beta}}

\newcommand{\delo}{\delta_{o}}
\newcommand{\delm}{\delta_{\mu}}
\newcommand{\go}{g_o}
\newcommand{\gm}{g_\mu}

\newcommand{\wco}{\omega_{co}}
\newcommand{\wcm}{\omega_{c\mu}}
\newcommand{\daok}{\delta_{ao,k}}
\newcommand{\damk}{\delta_{a\mu,k}}
\newcommand{\dao}{\delta_{ao}}
\newcommand{\dam}{\delta_{a\mu}}
\newcommand{\wo}{\omega_{o}}

\newcommand{\gok}{g_{o,k}}
\newcommand{\gmk}{g_{\mu,k}}
\newcommand{\gokc}{g_{o,k}^*}
\newcommand{\gmkc}{g_{\mu,k}^*}
\newcommand{\Wk}{\Omega_k}
\newcommand{\sig}[1]{\sigma_{#1}}
\newcommand{\sigk}[1]{\sigma_{#1,k}}
\newcommand{\w}[1]{\omega_{#1}}
\newcommand{\wk}[1]{\omega_{#1,k}}

\newcommand{\gamoc}{\gamma_{oc}}
\newcommand{\gamoi}{\gamma_{oi}}
\newcommand{\gammc}{\gamma_{\mu c}}
\newcommand{\gammi}{\gamma_{\mu i}}
\newcommand{\conj}[1]{{{#1}^*}}

\newcommand{\LL}{\mathcal{L}}

\newcommand{\La}{\mathcal{L}_{\aa}}
\newcommand{\Lac}{\mathcal{L}_{\conj{\aa}}}
\newcommand{\Lb}{\mathcal{L}_{\bb}}
\newcommand{\Lbc}{\mathcal{L}_{\conj{\bb}}}

\renewcommand{\aa}{\alpha}
\newcommand{\bb}{\beta}
\title{Theory of Microwave-Optical Conversion Using Rare-Earth Ion Dopants}

\begin{document}
\author{Peter S. Barnett}\affiliation{Dodd-Walls Centre for Photonic and Quantum Technologies and the Department of Physics, University of Otago, Dunedin, New Zealand}
\author{Jevon J. Longdell}\email{jevon.longdell@otago.ac.nz} \affiliation{Dodd-Walls Centre for Photonic and Quantum Technologies and the Department of Physics, University of Otago, Dunedin, New Zealand}
\title{Theory of Microwave-Optical Conversion Using Rare-Earth Ion Dopants}

\begin{abstract}

We develop a theoretical description of a device for coherent conversion of microwave to optical photons. For the device, dopant ions in a crystal are used as three-level systems, and interact with the fields inside overlapping microwave and optical cavities.  We develop a model for the cavity fields interacting with an ensemble of ions, and model the ions using an open quantum systems approach, while accounting for the effect of inhomogeneous broadening. Numerical methods are developed to allow us to accurately simulate the device. We also further develop a simplified model, applicable in the case of small cavity fields which is relevant to quantum information applications. This simplified model is used to predict the maximum conversion efficiency of the device. We investigate the effect of various parameters, and predict that conversion efficiency of above 80\% should be possible with currently existing experimental setups inside a dilution refrigerator. 
\end{abstract}
\maketitle

\section{Introduction}
Superconducting qubits are one of the main qubit designs for quantum computation, and have two states which are separated by a microwave frequency. This allows them to be easily coupled and controlled with microwave photons, typically in the 5-10\,GHz range. The relatively low energy of the microwave photons means that when transmitting the quantum information encoded in them, at all but millikelvin temperatures, the signal is overwhelmed by the effects of thermal noise. Optical photons, however, have much higher energy and so are immune to these effects of thermal noise, and additionally can be sent via existing fiber optic networks. Therefore there is much interest in  being able to coherently convert between microwave and optical photons, without destroying the encoded quantum information. 

There are several different experimental approaches for achieving microwave to optical photon upconversion \cite{lambert_coherent_2020}. Each of these methods aims to coherently combine an input microwave photon carrying the quantum information, with one or more optical photons. Some of these methods include using $\chi^{(2)}$ non-linear materials \cite{boyd_nonlinear_2003,khan_optical_2007,strekalov_efficient_2009}; collective spin systems where the microwaves can excite a magnon mode \cite{osada_cavity_2016,haigh_triple-resonant_2016,osada_brillouin_2018,everts_microwave_2019}; clouds of ultracold Rydberg atoms which takes advantage of the large transition dipole moment of the Rydberg atoms  \cite{han_coherent_2018,covey_microwave--optical_2019}; and optomechanical systems. The optomechanical approaches  have achieved the highest conversion efficiency at 47\% \cite{higginbotham_harnessing_2018}. This high efficiency optomechanical approach used a low frequency high Q mechanical resonator which led to a low bandwidth and added noise. Recently, a different optomechanical upconversion method has been demonstrated which appears to overcome these disadvantages~\cite{mirhosseini_quantum_2020}, in this work they achieved the milestone of generating optical photons from superconducting qubits but without the same high efficiency.


In this work we focus on using a crystal doped with rare-earth ions, which are used as three level systems. This scheme was originally proposed in \cite{williamson_magneto-optic_2014}, and has been investigated theoretically and experimentally in \cite{fernandez-gonzalvo_cavity-enhanced_2019}. Similar proposals for conversion using crystals doped with three level systems have also been made \cite{obrien_interfacing_2014,blum_interfacing_2015}. Treating the ions as three level systems, the lower two levels  ($\ket1$, $\ket2$) are the lowest Kramers doublet, which have been Zeeman split by an external magnetic field; and the third level $\ket3$ is an electronic excited state. The basic scheme is as follows (Figure \ref{fig:three_level_simple}): an input microwave photon excites the ion into the upper Zeeman level; a strong optical laser pump is applied to drive this further into the electronic excited state $\ket3$; the atom will then transfer back into the ground state, emitting an optical photon with the frequency equal to the sum of the two input photons. The device that we will investigate is composed of our doped crystal inside overlapping microwave and optical cavities. The optical cavity can be designed such that the process is phase matched, and there are two optical cavity modes which are close to resonant with the optical pump and the unconverted optical field. Upconversion using a similar atomic scheme is suggested in \cite{welinski_electron_2019}, but instead of using the microwave transition of the ground state doublet, and excited state doublet is used. 


 The model which we will develop is quite general, and should apply to devices using ensembles of any three level atoms. For our calculations we will consider erbium ions in yttrium orthosilicate. Erbium ions are of particular interest for microwave to optical upconversion because they  an optical transition around 1550\,nm, which is in the region where silica fibers have the lowest attenuation. Upconversion in such a device, using erbium, has been demonstrated experimentally \cite{fernandez-gonzalvo_cavity-enhanced_2019} with a maximum conversion efficiency of $1.3\times10^{-5}$.

We begin by using input--output formalism \cite{gardiner_quantum_2004,collett_squeezing_1984} to formulate equations for the optical and microwave cavity fields, this results in equations which are coupled by the interaction between the cavity fields and the atoms. We will then develop a description for the dynamics of the ensemble of atoms, and their interaction with the light fields in the cavities using a master equation approach. From here we can numerically solve for the cavity field equations, and in the case where we have no input optical field, the optical output field will be due entirely to upconverted microwave photons. 

We develop and test a linearised model which is applicable when the cavity inputs are small, such as the regime used for quantum information processing. We use this model to find experimental parameters to optimise the conversion efficiency. 

\section{The Theoretical Model}
\begin{figure}
    \centering
    \begin{subfigure}[t]{0.4\columnwidth}
    \includegraphics[width=\textwidth]{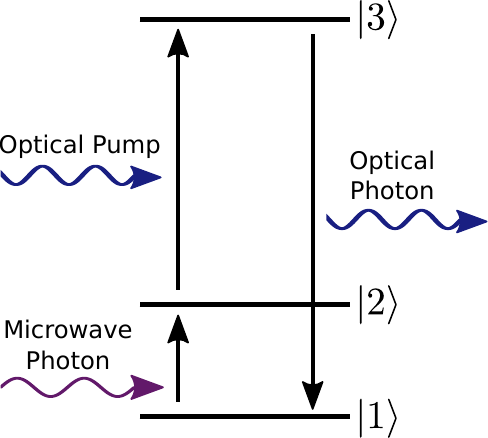}
    \caption{Energy level diagram}\label{fig:three_level_simple}
    \end{subfigure}
    \begin{subfigure}[t]{0.55\columnwidth}
    \includegraphics[width=\textwidth]{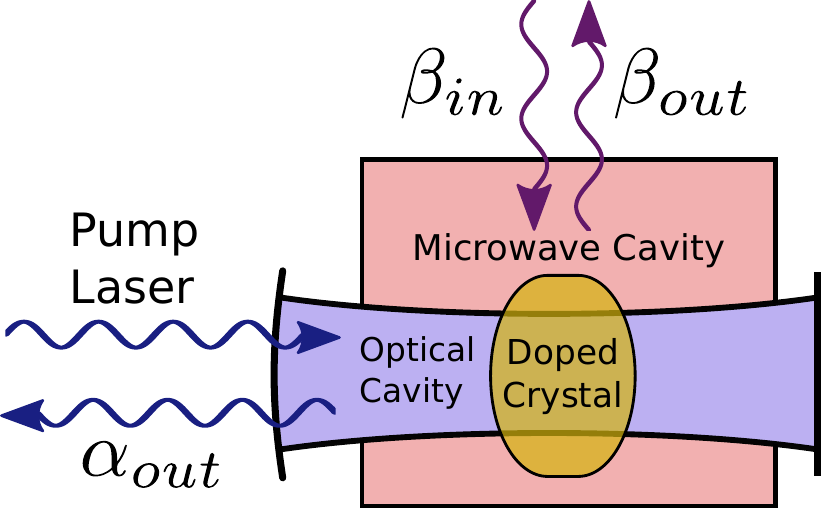}
    \caption{Double cavity upconversion device.}\label{fig:Double_cavity_device}
    \end{subfigure}
    \caption{(a) An energy level diagram of a three level atom absorbing a microwave photon and a photon from an optical pump, and emitting an optical photon. (b) The double cavity upconversion device, with a crystal doped with rare-earth ions in the overlapping modes of an optical and a microwave field. An input microwave photon ($\bb_{in}$) and a pump laser photon are combined via the interaction in (a) to produce an optical upconverted photon ($\aa_{out}$).}
\end{figure}

The principle of operation for the upconversion device is shown in Figure~\ref{fig:Double_cavity_device} \cite{williamson_magneto-optic_2014}.  

For the case of erbium ions, the $\ket1$ and $\ket2$ states are the Zeeman levels of \ce{^4I_{15/2}(Z_1)}, and $\ket3$ is one of the \ce{^4I_{13/2}(Y_1)} levels.


The Hamiltonian for the device can be expressed as 

\begin{equation}
\HH{}=\HH{fields}+\HH{atoms}+\HH{int}
\end{equation}
 where,
 \begin{align}
 \HH{fields}&=\wcm\bhd\bh+\wco\ahd\ah\\
 \HH{atoms} &=\sum_k \wk{12}\sigk{22}+ \wk{13}\sigk{33}\\
 \HH{int}	&=\sum_k \gmk \bh\sigk{21}+ \gok \ah \sigk{31}+\Wk \sigk{32} +\text{h.c.}
 \end{align}
 $\HH{fields}$ describes the energy of the cavity fields, where the frequency of the microwave (optical) cavity is $\wcm$ ($\wco$) and the corresponding annihilation operator is $\bh$ ($\ah$). 
 
 $\HH{atoms}$ is the energy of the atoms, and $\HH{int}$ is the Hamiltonian for the interaction between the cavity fields and the atoms. The sums represent the sums over all the atoms. For the $k$th atom, $\wk{nm}$ is the transition frequency between the $\ket{n}$ and $\ket{m}$ levels, $\sig{nn}$ is the population of $\ket{n}$, $\sigk{mn}$ is the atomic transition operator for the $\ket{m}\to\ket{n}$ transition, $\gmk$ ($\gok$) is the coupling between the microwave (optical) transition and the cavity field, and $\Wk$ is the Rabi frequency of the $\ket{2}\to\ket{3}$ transition driven by the pump laser.

\subsection{Cavity Fields}
 Using quantum input--output formalism from \cite{gardiner_quantum_2004,collett_squeezing_1984}, and following the initial working from \cite{fernandez-gonzalvo_cavity-enhanced_2019}, the equations of motion for the microwave and optical cavity field operators in the time domain are 
 \begin{align}\label{eq:a_time_domain_double_imput}
\diff{\ah}{t}&=-i(\wco\ah +\sum_k\gokc \sigk{13})-\frac{\gamoc+\gamoi}{2}\ah +\sqrt{\gamoc}\hat{a}_{in} +\sqrt{\gamoi}\hat{a}_{in,i}\\
\diff{\bh}{t}&=-i(\wcm\bh +\sum_k\gmkc \sigk{12})-\frac{\gammc+\gammi}{2}\bh +\sqrt{\gammc}\hat{b}_{in} +\sqrt{\gammi}\hat{b}_{in,i} \label{eq:b_time_domain_double_input}
\end{align}
where for the microwave (optical) cavity, $\gammc$ ($\gamoc$) and $\gammi$ ($\gamoi$) are the coupling and intrinsic losses, $\hat{b}_{in}$ ($\hat{a}_{in}$) represents the input field into the cavity via the port, and $\hat{b}_{in,i}$ ($\hat{a}_{in,i}$) represents any other input. 

We now make a semi-classical approximation, treating our fields as complex amplitudes rather than operators, $\hat{b}\to\bb, \ \hat{a}\to\aa$. Additionally $\sigk{12}$ and $\sigk{13}$, and the input fields will also be treated as complex numbers, and any undriven input fields are set to zero, $\bb_{in,i}=\aa_{in,i}=0$. 

 \begin{align} 
 \dot{\bb}(t)=-i(\wcm \bb+\sum_k\gmkc\sigk{12})-\frac{\gammc+\gammi}{2}\bb+\sqrt{\gammc}\bb_{in}\\
 \dot{\aa}(t)=-i(\wco \aa+\sum_k\gokc\sigk{13})-\frac{\gamoc+\gamoi}{2}\aa+\sqrt{\gamoc}\aa_{in} 
 \end{align}
 
 Taking the Fourier transform of these relations gives,
 \begin{align} \label{eq:b_delm}
\bt(\delm)&=\frac{-i\sum_k\gmkc\sigk{12}}{(\gammc+\gammi)/2-i\delm}+\frac{\sqrt{\gammc}\bt_{in}}{(\gammc+\gammi)/2-i\delm}\\
\at(\delm)&=\frac{-i\sum_k\gokc\sigk{13}}{(\gamoc+\gamoi)/2-i\delo}+\frac{\sqrt{\gamoc}\at_{in}}{(\gamoc+\gamoi)/2-i\delo} \label{eq:a_delo}
 \end{align}
 which tell us the cavity field amplitudes as a function of the detunings between input fields and the cavities,
 \begin{equation} \label{eq:delm_delo}
 \delm=\wo-\wcm, \qquad \delo=\wo-\wco
 \end{equation}
 These equations for the cavity fields are coupled and nonlinear because of the interactions between the light fields and the atoms, these interactions appear in the equations as sums over the atomic transition operators $\sigk{12}$ and $\sigk{13}$.
 The values used for the atomic transition operators for the individual atoms will depend on the frequencies of the atomic transitions and the cavity fields, as well as the cavity field amplitudes $\bt$ and $\at$. 
 
 Equations \ref{eq:b_delm} and \ref{eq:a_delo} will be solved to yield the cavity field amplitudes, and from the input--output theory \cite{gardiner_quantum_2004} the output fields are given by
 \begin{equation}\label{eq:bout_aout}
 \bb_{out}=\sqrt{\gammc}\bb,\qquad \aa_{out}=\sqrt{\gamoc}\aa 
 \end{equation}
 
 Therefore finding the optical output field $\aa_{out}$ as a function of the input microwave field $\bb_{in}$ will allow is to find the conversion efficiency of the device. 
\subsection{Atomic Dynamics} 

\begin{figure}[ht]
	\centering
	\includegraphics[]{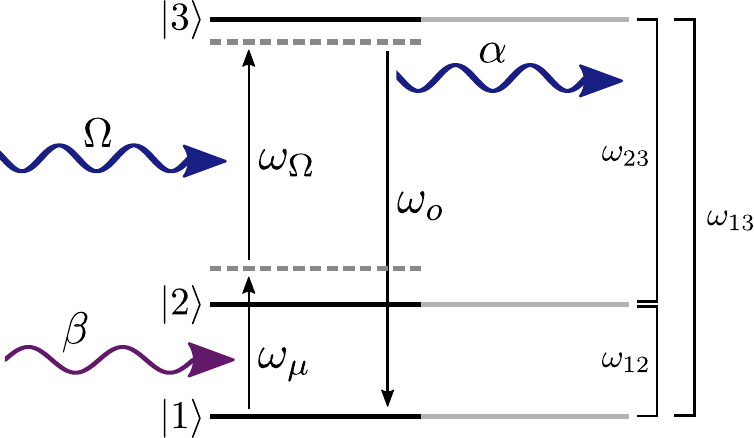}
	\caption{Energy level diagram for the scheme using the ground state microwave transition, showing frequencies of the atomic transitions and the light fields. The fields may be detuned from the atomic transitions. The microwave field is represented by $\bb$, the upconverted optical field is represented by $\aa$, and the Rabi frequency for the pump field is $\Omega$.  }
	\label{fig:3lvl_detuned}
\end{figure}

To solve for the cavity fields $\bb$ and $\aa$, we need to find the classical values for the atomic transition operator terms $\sig{12}$ and $\sig{13}$ for each individual atom. We initially consider a three level atom interacting with microwave and optical fields (with classical amplitudes $\bb$ and $\aa$ respectively), and with the $\ket2\to\ket3$ transition driven by a coherent driving field. The three level atoms are here modeled in a similar fashion as in \cite{fernandez-gonzalvo_cavity-enhanced_2019}.

To find $\sigk{13}$ and $\sigk{12}$ we first find the steady state density matrix, because the individual elements correspond to the ensemble average of the classical values of the atomic transition operators,
\begin{equation}
    \rho_{nm}=\sum_j p_j \braket{\sig{nm}}_j
\end{equation}
where $p_j$ is the statistical probability that the the atom is in state $\ket{\psi_j}$.

The steady state density matrix is found by solving
\begin{equation}\label{eq:rho_steady_state}
\dot{\rho}=\LL \rho=0
\end{equation} 
where $\LL$ is the Liouvillian superoperator which accounts for both Hamiltonian evolution and damping processes, and can be written as
\begin{equation}
\LL\rho=-\frac{i}{\hbar}\left[\HH{},\rho\right]+\Gamma\rho
\end{equation}
Where $\HH{}$ here is the Hamiltonian for a single three level atom interacting with the various light fields. This is in the frame rotating with the light fields because otherwise we would not be able to find a steady state $\rho$.
$\HH{}$ can be expressed in matrix form as
\begin{equation}\label{eq:H_atom_matrix}
\HH{}=\begin{bmatrix}
0 & \gm\conj{\bb} & \go\conj{\aa} \\
\gm \bb & \dam-\delm & \conj{\Omega}\\
\go \aa & \Omega & \dao-\delo
\end{bmatrix}
\end{equation}
Here the energy of the levels in this rotating frame are expressed in terms of detunings rather than the absolute frequencies;
$\dam$ and $\dao$ represent the detuning between the cavity and atomic transition for this specific atom,
\begin{equation}
\dam=\w{12}-\wcm,\qquad \dao=\w{13}-\wco,
\end{equation} 
$\delm$ and $\delo$ are the detunings between the fields and the cavities (Equation \ref{eq:delm_delo}). These detunings are used because the atomic detunings ($\dam$ and $\dao$) vary for each atom, and later we will integrate over these detunings while holding the others constant.

The superoperator $\Gamma$ describes the effects of damping, and may be expressed as $\Gamma\rho=\Gamma_{12}\rho+\Gamma_{23}\rho+\Gamma_{13}\rho+\Gamma_{2d}\rho+\Gamma_{3d}\rho$, where $\Gamma_{nm}$ results in population decay from $\ket{m}$ to $\ket{n}$, and $\Gamma_{nd}$ results in additional dephasing of state $\ket{n}$ 
\begin{align}
\Gamma_{nm}\rho&=\frac{\gamma_{nm}}{2}(N_{nm}+1)(2\sig{nm}\rho\sig{nm}-\sig{mn}\sig{nm}\rho-\rho\sig{mn}\sig{nm})\\
			   &+\frac{\gamma_{nm}}{2}N_{nm}(2\sig{mn}\rho\sig{mn}-\sig{nm}\sig{mn}\rho-\rho\sig{nm}\sig{mn})\nonumber\\
\Gamma_{nd}\rho&=\frac{\gamma_{nd}}{2}(2\sig{nn}\rho\sig{nn}-\sig{nn}\rho-\rho\sig{nn})
\end{align}
$\gamma_{nm}$ is the damping rate for the $\ket{m}\to\ket{n}$ transition, $\gamma_{nd}$ is the dephasing rate of state $\ket{n}$, and $N_{nm}$ is the excitation number for the transition at the temperature of the atom. $N_{nm}$ is given by the Planck distribution, and will only be non-zero for the microwave $\ket1\to\ket2$ transition because the transition frequencies of the optical transitions are so high. 

We are now in a position to solve for the steady state density matrix (Equation \ref{eq:rho_steady_state}). To do this we use the method described in \cite{fernandez-gonzalvo_cavity-enhanced_2019} and express $\rho$ as a $9\times1$ vector, and $\LL$ as a $9\times9$ matrix. By taking into account the normalisation condition on $\rho$ that $\rho_{11}+\rho_{22}+\rho_{33}=1$, we can invert $\LL$ and solve for the unique steady state density matrix solution. This allows us to find $\rho$, and thus the classical values of the atomic transition operators $\sigk{13}$ and $\sigk{12}$, as a function of the various detunings and the cavity field amplitudes $\aa$ and $\bb$. 

The atomic terms that appear in our cavity field equations are sums over the atomic transition operators for each of the individual atoms.

In the doped crystal, each atom will be in a slightly different environment and hence there will be inhomogeneous broadening of the spectral lines. This means that each atom will have different transition frequencies and different atomic detunings.  
  Because the number of atoms is so large $(\sim10^{16}\text{ atoms})$ we are unable to perform these sums directly, and instead we approximate the sums as integrals. 
We assume the microwave and optical atomic detunings are normally distributed around a central frequency, and so integrate the density matrix terms over the detunings, with weightings given by the normal distributions. The sum terms can then be expressed as,
\begin{align}
&\sum_k \gokc\sigk{13}(\daok,\damk)=\nonumber\\\label{eq:rho13_integral}
&N_o\go\int d\dao \int d\dam G_o(\dao)G_\mu(\dam)\rho_{13}(\dao,\dam)\\
&\sum_k \gmkc\sigk{12}(\daok,\damk)=\nonumber\\\label{eq:rho12_integral}
&N_\mu\gm\int d\dao \int d\dam G_o(\dao)G_\mu(\dam)\rho_{12}(\dao,\dam)
\end{align}
where  $G_o(\dao)$ and $G_\mu(\dam)$ are normal distributions representing the optical and microwave inhomogeneous distributions. These two distributions will be centered around the mean optical and microwave atomic frequencies, and the width will be determined by the inhomogeneous linewidth.  We are assuming that the coupling strengths $\go$ and $\gm$ are real, and the same for all the atoms. As well as being a function of the detunings, the density matrix elements will also depend on the cavity field amplitudes. 

\section{Numerical Methods}

\begin{figure}
	\centering
	\begin{subfigure}[b]{1\columnwidth}
		\includegraphics[width=\textwidth]{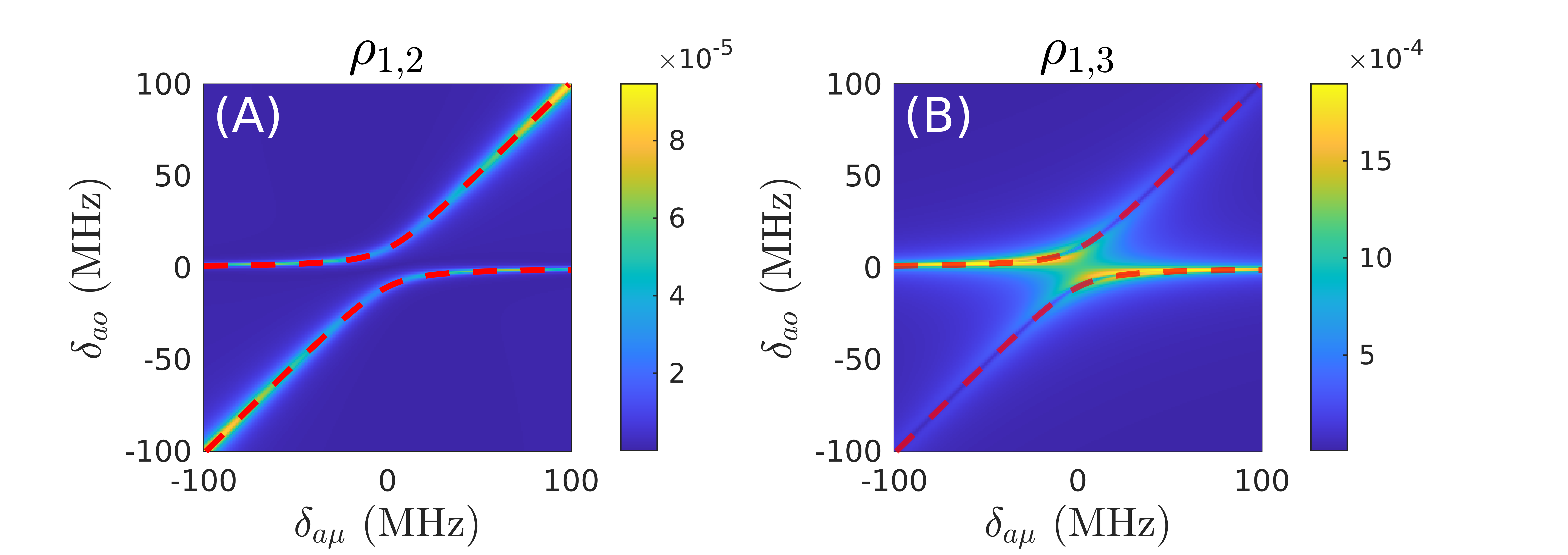}
		\caption{Large microwave field}\label{fig:rho_ds_bigb}
	\end{subfigure}
	\\
	\begin{subfigure}[b]{1\columnwidth}
	\includegraphics[width=\textwidth]{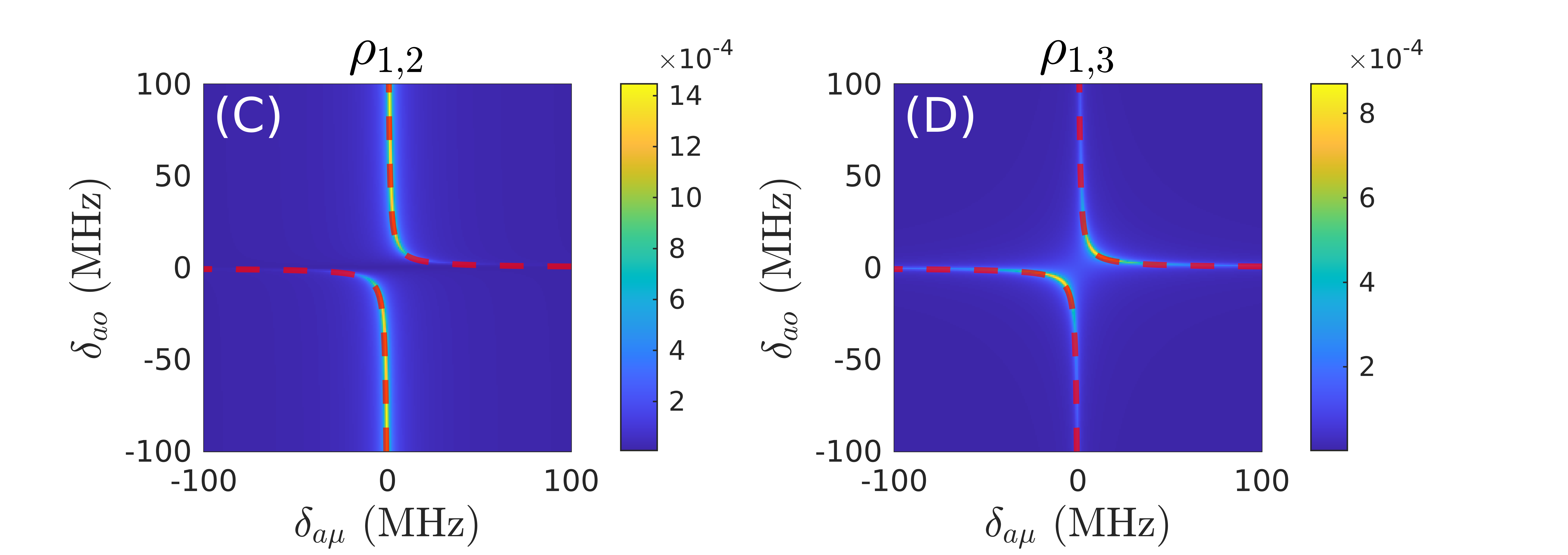}
	\caption{Small microwave field}\label{fig:rho_ds_smallb}
\end{subfigure}
\caption{The single atom coherences $\rho_{12}$ and $\rho_{13}$ as functions of the atomic detunings, using parameters from Table \ref{tab:parameters}, for large microwave and small pump field, and small microwave and large pump field. The red lines indicate the detunings corresponding to the degenerate dressed states. }
	\label{fig:rho_ds}
\end{figure}

 The components of $\rho$ can change rapidly with frequency around regions where the detunings $\dao$ and $\dam$ correspond to degenerate eigenvalues of the atomic Hamiltonian (Equation \ref{eq:H_atom_matrix}). This can be seen in Figure~\ref{fig:rho_ds}. The eigenvalues of Equation~\ref{eq:H_atom_matrix} correspond to the energy of the dressed states of a single atom interacting with the fields. As we are are working in the rotating frame where both the microwave and optical fields have zero frequency,  these eigenvalues being degenerate corresponds to the fields directly driving the transitions between the dressed states.

When we numerically evaluate the integrals in Equations \ref{eq:rho13_integral} and \ref{eq:rho12_integral} it is important to know locations of these dressed states so that they are not missed. This is particularly important because the homogeneous linewidth is much smaller than the inhomogeneous linewidth.

 For each value of the atomic optical detuning $\dao$  there will be an atomic microwave detuning $\dam$ which corresponds to the dressed state transition. For the numerical integration we will treat the integral over $\dao$ as the outer integral, and so for each value of $\dao$ we will find the value of $\dam$ over which to split up the $\dam$ integral. We do this by numerical root finding using an analytical solution to a simpler problem as an initial guess.

\begin{figure}
	\centering
	\includegraphics[width=0.3\textwidth]{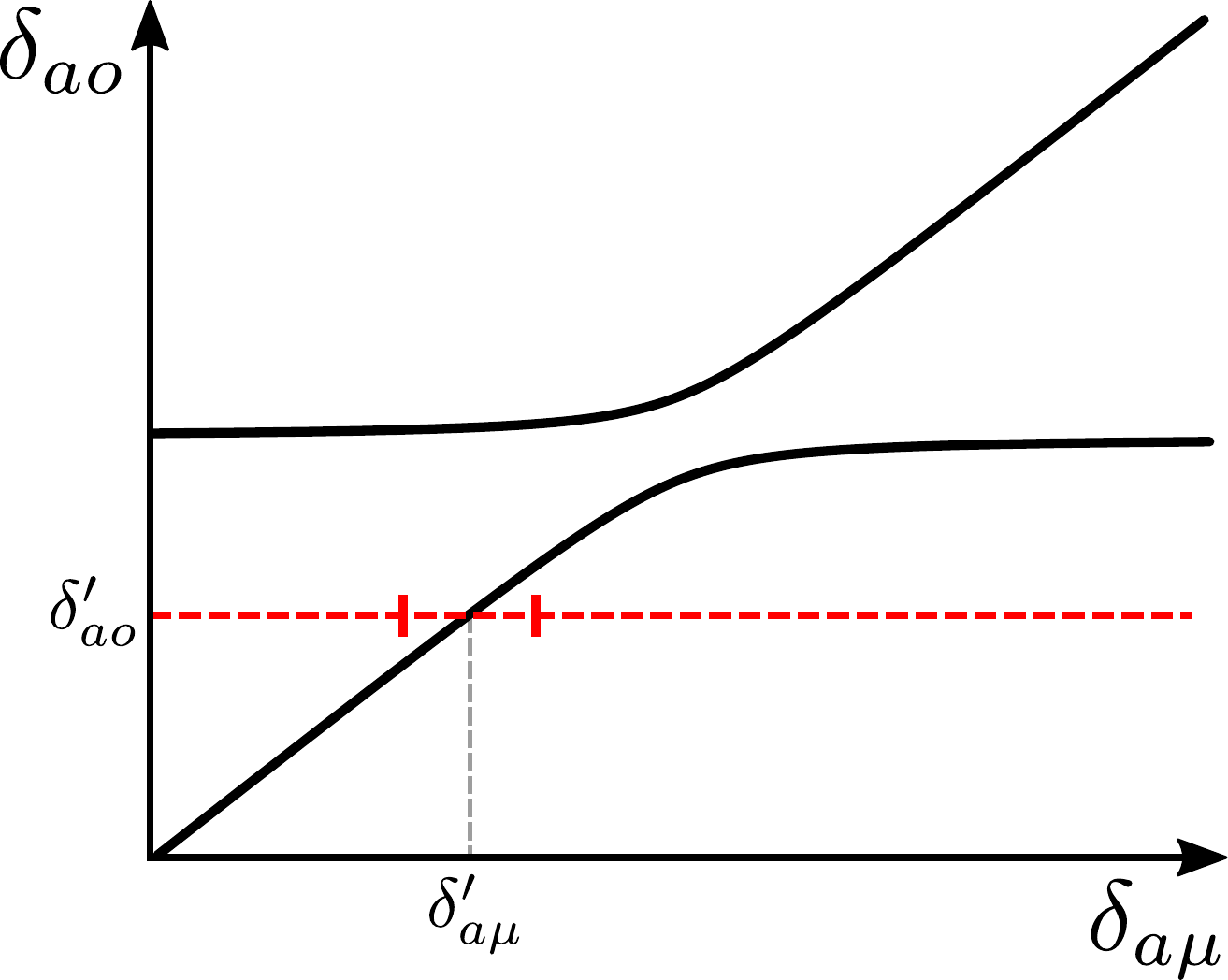}
	\caption{Schematic demonstrating how the integration regions are split up. The black curve represents the centre of the peak of $\rho(\dao,\dam)$. When evaluating the outer integral over $\dao$, for each value of the integration point $\dao'$ there is a $\dam'$ corresponding to the peak. So when we evaluate the inner integral over $\dam$, integrating along the dashed red line, we will split this integral up around $\dam'$. This will be done for every value of $\dao'$. }
	\label{fig:int_fig}
\end{figure}

\subsection{Method for finding detunings for degenerate dressed states}
The eigenvalues $\lambda$ of our Hamiltonian are given by 
\begin{equation}
\det(\HH{}-\mathbbm{1}\lambda)=0
\end{equation}
Because $\HH{}$ is a $3\times3$ matrix, this determinant will be a cubic polynomial in $\lambda$. For a cubic, if the discriminant is zero then the polynomial has at least 2 repeated roots. This is exactly what we are looking for, because repeated roots mean that the Hamiltonian has degenerate eigenvalues. Therefore we want to find detunings $\dao$ and $\dam$ such that 
\begin{equation} \label{eq:disc_det_H}
\text{Disc}_\lambda(\det(\HH{}(\dao,\dam)-\mathbbm{1}\lambda))=0 
\end{equation} 
For a value of the optical detuning $\dao$, we want to find which value of $\dam$ satisfies Equation \ref{eq:disc_det_H}. This can be done numerically, for which we need an initial guess of the value of $\dam$.

For these initial guesses we will assume that the optical output field $\aa$ is negligible, and then work in the regime where we can either ignore the effect of the microwave field or the optical pump laser. 

In the case where the optical pump is small and can be ignored, $\Omega\to0$, and our Hamiltonian becomes 
\begin{equation}
\HH{}=\begin{bmatrix}
0 & \gm\conj{\bb} & 0 \\
\gm \bb & \dam-\delm & 0\\
0 & 0& \dao-\delo
\end{bmatrix}
\end{equation}

And so the degenerate dressed state detunings are 
\begin{equation}
\dam=-\frac{|\gm \bb|^2}{\dao-\delo}+\dao-\delo+\delm
\end{equation}

So for a given optical detuning $\dao$, $\dam$ is the microwave detuning which makes the dressed states degenerate. 
Alternatively, when the microwave field is negligible, 
\begin{equation}
\HH{}=\begin{bmatrix}
0&0& 0\\
0 & \dam-\delm & \conj{\Omega}\\
0 & \Omega & \dao-\delo
\end{bmatrix}
\end{equation}
where the degenerate dressed state detunings are given by 
\begin{equation}
\dam=\frac{|\Omega|^2}{\dao-\delo}+\delm
\end{equation}
The magnitudes of $\rho_{12}$ and $\rho_{13}$ are plotted in Figure \ref{fig:rho_ds} as functions of the microwave and optical atomic detunings, and the detunings corresponding to the degenerate dressed states are where these functions are rapidly changing. The atomic parameters for Figure \ref{fig:rho_ds} are the same as are used in Section \ref{sec:Optimising}; for \ref{fig:rho_ds_bigb} the microwave field corresponds to 5\,dBm of input microwave power, and the pump laser power is 1\,pW; for \ref{fig:rho_ds_smallb} the microwave power is -75\,dBm and the pump power is 100\,mW.

For performing the integrals in Equations \ref{eq:rho13_integral} and \ref{eq:rho12_integral}, we treat the integral over the optical detuning $\dao$ as the outer integral. When performing this integral numerically, for each value of $\dao$ we have to perform the integral over $\dam$ (Figure~\ref{fig:int_fig}). For a given value of $\dao$, the inner integral over $\dam$ will be peaked when the detunings mean that the dressed states are degenerate. To take these peaks into account, Equation \ref{eq:disc_det_H} can be used to find the value of $\dam$ where the peak occurs, for the given value of $\dao$, and then the integral will be split around this value. 

 Finally, having found appropriate bounds for our numerical integrals, each individual section is integrated using Gauss-Lobatto quadrature, a quadrature integration method which uses the bounds of integration as 2 of the grid points and will ensure that integration points doesn't miss the peaks of the sharply peaked functions. These integrated terms are the density matrix terms integrated over the inhomogeneous broadening distribution, and can be used to find the atomic transition operator sum terms for our classical cavity field equations.

This whole procedure allows us to solve for the atomic transition operator sum terms in the cavity field equations, as functions of the cavity fields. These atomic transition operator terms depend nonlinearly on the cavity fields, and so we will numerically solve the cavity field equations for $\aa$ and $\bb$, as a function of the microwave and optical detuning.

\section{Linear Approximation}\label{sec:linear}
It is very slow to evaluate the cavity fields using an iterative numerical method because this requires us to evaluate the atomic ensemble terms many times. This would make optimising the device conversion efficiency also very slow, and so in this section we develop an approximate form of the cavity field equations which does not require iterative methods to solve and so is much faster. 

The atomic terms in Equations \ref{eq:b_delm} and \ref{eq:a_delo} depend nonlinearly on the cavity field amplitudes, which means that the equations must be solved numerically. However, if the cavity fields are very small we can make the approximation that the atomic terms only depend linearly on the cavity fields allowing us to solve for the cavity fields directly. It should be pointed out that we are only considering the fields in the microwave and optical signal cavities to be small. We assume nothing about the strength of the optical pump field.

We start by noting that, without approximation, $\LL$ can be separated into terms which each depend only linearly on one of the cavity fields or their conjugates,
\begin{equation}
\LL=\LL_0+\aa\La+\conj{\aa}\Lac +b\Lb+\conj{\bb}\Lbc 
\end{equation}
$\LL_0$ corresponds to the case where $\bb=\aa=0$, but there may still be the strong pump laser and excitation by thermal photons. We now make the approximation that our density matrix depends only linearly on the cavity fields,
\begin{equation}
\rho\approx\rho_0+\aa\rho_\aa+\conj{\aa}\rho_\conj{\aa}+\bb\rho_\bb+\conj{\bb}\rho_\conj{\bb}
\end{equation}
We are solving for steady state, $\dot{\rho}=\LL\rho=0$, and so 
\begin{equation}\label{eq:Lrho1}
(\LL_0+\aa\La+... )(\rho_0+\aa\rho_\aa+...
)=0
\end{equation}
$\LL_0\rho_0=0$ because this is steady state master equation  with $\bb=\aa=0$. If we assume that terms that are the product of two fields are negligible, Equation \ref{eq:Lrho1} can be rearranged to give
\begin{align} 
0&=\aa(\LL_0\rho_\aa+\La\rho_0)+\conj{\aa}(\LL_0\rho_\conj{\aa}+\Lac\rho_0)\\
&+\bb(\LL_0\rho_\bb+\Lb\rho_0)+\conj{\bb}(\LL_0\rho_\conj{\bb}+\Lbc\rho_0)
\end{align}
Because the field amplitudes and their conjugates are treated as independent variables each of these terms must individually be zero, $\LL_0\rho_x+\LL_x\rho_0=0$ for $x=\aa,\conj{\aa},\bb,\conj{\bb}$. And so we can find each $\rho_x$,
\begin{equation}\label{eq:rhox}
	\rho_x=-\LL_0^{-1}\LL_x\rho_0
\end{equation}

For this linear approximation we will replace the atomic terms in the cavity field equations (Equations \ref{eq:b_delm} and \ref{eq:a_delo}) $\sigk{12}$ and $\sigk{13}$ with the relevant elements of the linearised $\rho$. For this approximation, only $\rho_\aa$ and $\rho_\bb$ contribute nonzero elements,
\begin{align}
\sig{12}\approx\aa\rho_{\aa,12}+\bb\rho_{\bb,12}\\
\sig{13}\approx\aa\rho_{\aa,13}+\bb\rho_{\bb,13}
\end{align}

and hence the atomic ensemble terms can be written as 
\begin{align}
S_{12}&=\aa\sum_k\gmk \rho_{\aa,12,k}+ \bb\sum_k\gmk \rho_{\bb,12,k}\\
	  &=\aa S_{\aa,12}+\bb S_{\bb,12}	
\end{align}
and likewise,
\begin{equation}
S_{13}=\aa S_{\aa,13}+\bb S_{\bb,13}	
\end{equation}

We can calculate the linearised atomic ensemble terms by first calculating $\rho_0$, from this we can find the single atom $\rho_\aa$ and $\rho_\bb$ using Equation \ref{eq:rhox}. We then integrate the single atom coherences over the inhomogeneous distribution as we did in the unlinearised case to give us the linearised atomic ensemble terms. These terms do not depend on the cavity fields amplitudes $\aa$ and $\bb$, but will depend on the frequencies of the cavity fields, as well as distribution of the atomic transition frequencies. 

\begin{equation}
S_{\aa,12}=\int d\dao d\dam G(\dao,\dam) \rho_{\aa,12}(\dao,\dam)
\end{equation}
This yields equations for the cavity fields which only depend linearly on $\bb$ and $\aa$,
\begin{align}
-i \delm b &=-i (\aa S_{\aa,12}+\bb S_{\bb,12})-\frac{\gammc+\gammi}{2}\bb+\sqrt{\gammc}\bb_{in}\\
-i \delo a &=-i (\aa S_{\aa,13}+\bb S_{\bb,13})-\frac{\gamoc+\gamoi}{2}\aa+\sqrt{\gamoc}\aa_{in}
\end{align}

These equations are linear and so can be solved analytically. To do this we express them as a matrix
\begin{equation}
\begin{bmatrix}
\sqrt{\gammc}\bb_{in}\\[6pt] 
\sqrt{\gamoc}\aa_{in}
\end{bmatrix}=
\begin{bmatrix}
i S_{\aa,12} & i S_{\bb,12}-i\delm +\frac{\gammc+\gammi}{2}\\[6pt]
i S_{\aa,13}-i\delo +\frac{\gamoc +\gamoi}{2} & i S_{\bb,13}
\end{bmatrix}
\begin{bmatrix}
\bb\\[6pt]
\aa
\end{bmatrix}
\end{equation}

Using the relation between the output fields and the cavity fields (Equation \ref{eq:bout_aout}), we can now express the output fields in terms of the two input fields,
\begin{align}
\aa_{out}=C_{\aa\aa}\aa_{in}+C_{\aa\bb}\bb_{in}\\
\bb_{out}=C_{\bb\aa}\aa_{in}+C_{\bb\bb}\bb_{in}
\end{align}
where
\begin{widetext}
\begin{align}
C_{\aa\aa}=\frac{\gamoc\left(iS_{\bb,12}-i\delm+\frac{\gammc+\gammi}{2}\right)}{S_{\aa,12}S_{\bb,13}+\left(iS_{\aa,13}-i\delo+\frac{\gamoc+\gamoi}{2}\right)\left(iS_{\bb,12}-i\delm+\frac{\gammc+\gammi}{2}\right)}\label{eq:Caa_linear}\\
C_{\aa\bb}=-\frac{iS_{\bb,13}\sqrt{\gammc\gamoc}}{S_{\aa,12}S_{\bb,13}+\left(iS_{\aa,13}-i\delo+\frac{\gamoc+\gamoi}{2}\right)\left(iS_{\bb,12}-i\delm+\frac{\gammc+\gammi}{2}\right)} \label{eq:Cab_linear}\\ 
C_{\bb\aa}=-\frac{iS_{\aa,12}\sqrt{\gammc\gamoc}}{S_{\aa,12}S_{\bb,13}+\left(iS_{\aa,13}-i\delo+\frac{\gamoc+\gamoi}{2}\right)\left(iS_{\bb,12}-i\delm+\frac{\gammc+\gammi}{2}\right)}\label{eq:Cba_linear}\\ 
C_{\bb\bb}=\frac{\gammc\left(iS_{\aa,13}-i\delo+\frac{\gamoc+\gamoi}{2}\right)}{S_{\aa,12}S_{\bb,13}+\left(iS_{\aa,13}-i\delo+\frac{\gamoc+\gamoi}{2}\right)\left(iS_{\bb,12}-i\delm+\frac{\gammc+\gammi}{2}\right)}\label{eq:Cbb_linear}
\end{align}
$C_{\aa\aa}$ and $C_{\bb\bb}$ give us the transmission through the optical and microwave cavities,
\begin{equation}
\left|\frac{\aa_{out}}{\aa_{in}}\right|^2=|C_{\aa\aa}|^2,\qquad\left|\frac{\bb_{out}}{\bb_{in}}\right|^2=|C_{\bb\bb}|^2
\end{equation}
$C_{\aa\bb}$ and $C_{\bb\aa}$ give the conversion efficiency from microwave photons to optical photons and vice versa,
\begin{equation}
\left|\frac{\aa_{out}}{\bb_{in}}\right|^2=|C_{\aa\bb}|^2,\qquad\left|\frac{\bb_{out}}{\aa_{in}}\right|^2=|C_{\bb\aa}|^2
\end{equation}
And hence the conversion efficiency which we want to optimise is given by 
\begin{equation}\label{eq:Cab_sq}
|C_{\aa\bb}|^2=\left|\frac{S_{\bb,13}\sqrt{\gammc\gamoc}}{S_{\aa,12}S_{\bb,13}+\left(iS_{\aa,13}-i\delo+\frac{\gamoc+\gamoi}{2}\right)\left(iS_{\bb,12}-i\delm+\frac{\gammc+\gammi}{2}\right)}\right|^2
\end{equation}

This equation is very fast to evaluate, the atomic ensemble terms only need be evaluated once for each value of the detunings, rather than many times as they would be using an iterative method. 
\end{widetext}

\section{Optimising Upconversion Efficiency}\label{sec:Optimising}
The regime relevant to quantum information processing is expected to be close to the single photon regime. This means that the microwave and optical signal fields will be very small, and we expect saturation effects to be negligible and so we are able to use the linear approximation model (Section \ref{sec:linear}).  

The parameters which characterise the \ce{Er{{:}}Y_2SiO_5} experiments and are used for the simulations are summarised in Table \ref{tab:parameters}. These parameters come from a combination of recent experimental measurements \cite{gavin_in_prep} performed at low temperature in a dilution refrigerator and from previous literature.

\begin{table}
    
	\centering
	\begin{tabular}{lc}
		\hline \hline
		Parameter & Value \\ [0.5ex]
		\hline
        $1\leftrightarrow 3$  dipole moment		$d_{13}$ $^\ddagger$ & $1.63\times10^{-32}$\,Cm\\
	    $2\leftrightarrow 3$  dipole moment		$d_{23}$ $^\ddagger$& $1.15\times10^{-32}$\,Cm\\
		Lifetime of $\ket3$ $\tau_{3}$ $^\ddagger$& 11\,ms\\
		Lifetime of $\ket2$ $\tau_{2}$ $^\dagger$& 11\,s  \\
		Optical inhomogeneous linewidth $\sigma_o$ $^\dagger$& $2\pi\cdot419$\,MHz \\
		Microwave inhomogeneous linewidth $\sigma_\mu$ $^\dagger$& $2\pi\cdot5$\,MHz \\
		Optical coupling strength $g_o$ $^\ddagger$& 51.9\,Hz \\
		Microwave coupling strength $g_\mu$ $^\ddagger$& 1.04\,Hz \\
		Microwave cavity intrinsic loss $\gammi$ $^\dagger$& $2\pi\cdot650$\,kHz\\
		Microwave cavity coupling loss $\gammc$ $^\dagger$& $2\pi\cdot1.5$\,MHz\\
		Optical cavity intrinsic loss $\gamoi$ $^\ddagger$& $2\pi\cdot7.95$\,MHz\\
		Optical cavity coupling loss $\gamoc$ $^\ddagger$& $2\pi\cdot1.7$\,MHz\\
		\hline \hline
		$^\dagger$ From measurements. \\
		$^\ddagger$ From literature \cite{fernandez-gonzalvo_cavity-enhanced_2019, bottger_spectroscopy_2006}.
	\end{tabular}
\caption{\label{tab:parameters} The parameters characterising the atomic properties of erbium and the cavities which were used for the calculations, unless otherwise specified.}
\end{table}

To maximise the conversion efficiency we want to have the intra cavity fields as large as possible. If there was no coupling between the atoms and the cavity fields, then the field amplitudes would be largest when the the fields were resonant with the cavity. However, the interaction between the cavity fields and the atoms will `pull' the cavity resonance and the cavity fields will be largest around the dressed states of the atoms and the cavity. Approximating the atoms as identical two level atoms, the cavity detunings for the given atomic detunings are
\begin{subequations}\label{eq:atom_cavity_ds}
\begin{align}
\delta_{c\mu}=\frac{N_\mu \gm^2}{\dam}\\
\delta_{co}=\frac{N_o\go^2}{\dao}
\end{align}
\end{subequations}
Where $N_o$ is the number of atoms driven by the optical field, and $N_\mu$ is the effective number of atoms for the microwave transition, which will be influenced by the temperature $N_\mu=(\rho_{11}-\rho_{22})N=\frac{\exp(\hbar\omega/k_bT)-1}{\exp(\hbar\omega/k_bT)+1}N$, where $N$ is the total number of atoms.

\begin{figure}
	\centering
	\begin{subfigure}[b]{1\columnwidth}
		\includegraphics[width=\textwidth]{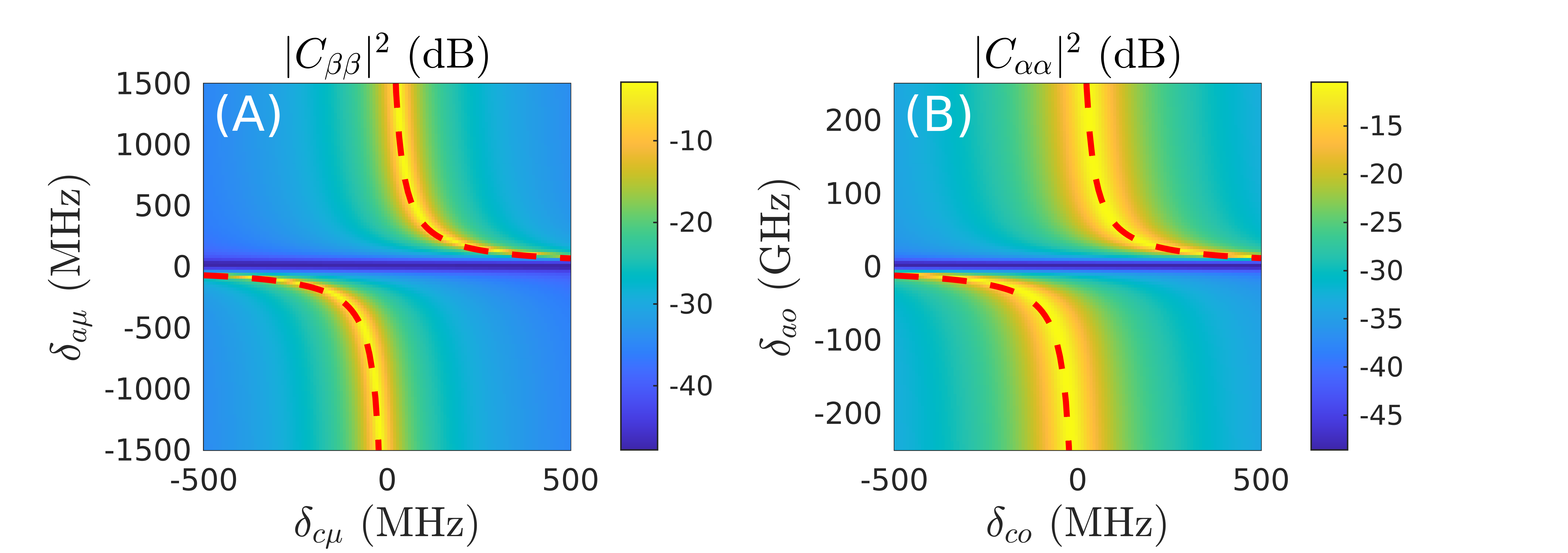}
		\caption{Cavity transmission}
	\end{subfigure}
	\\
	\begin{subfigure}[b]{1\columnwidth}
		\includegraphics[width=\textwidth]{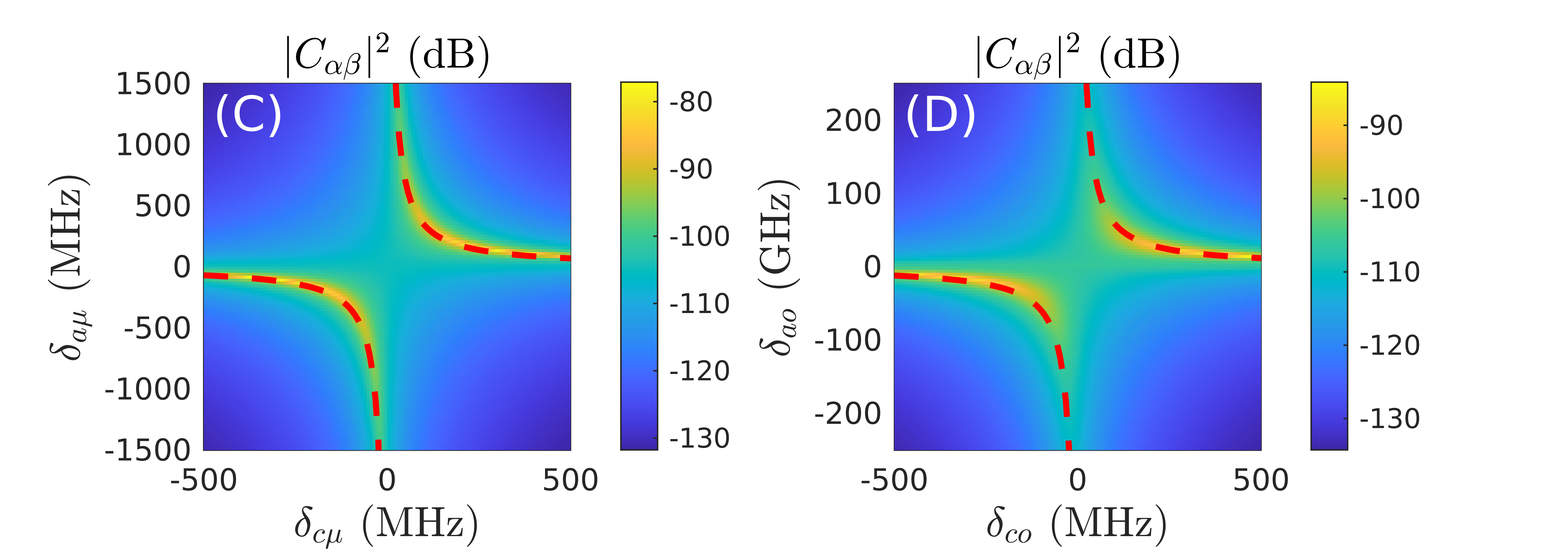}
		\caption{Conversion efficiency}
	\end{subfigure}
	\caption{Cavity transmission and conversion efficiency as functions of both the cavity and atomic detunings (microwave and optical). The red lines correspond to the approximate dressed states of the atoms and cavity (Equation \ref{eq:atom_cavity_ds}). For this simulation there are $N=4.8\times10^{16}$ atoms, with $N_o=2.2\times10^{15}$ in the optical beam path, at a temperature of 100\,mK.}    \label{fig:cavity_shift}
\end{figure}

Initially to improve the conversion efficiency we can scan the microwave and optical atomic detunings, and see how the conversion efficiency (Equation \ref{eq:Cab_sq}) changes as shown in Figure \ref{fig:cavity_shift}. To keep the cavity fields large, the cavity detunings were set to the dressed state detunings given by Equations \ref{eq:atom_cavity_ds} and so these cavity detunings will change as functions of the atomic detunings. 

As well as having control over the atomic and cavity detunings, experimentally we also have some control over the coupling rates $\gamoc$ and $\gammc$ for the optical and microwave cavities, as well as control over the strength of the optical pump laser. When using a Fabry-P\'eriot resonator as the optical cavity, we have control over $\gamoc$ by changing the reflectivity of the mirrors, and when using a whispering gallery mode resonator the coupling rate can be changed by changing the distance between the resonator and the coupling prism. Similarly we can change the microwave coupling by adjusting the antennae. 


We want to be able to maximise the conversion efficiency, which will be affected by several factors including the laser pump power, the intrinsic cavity losses and the temperature of the atoms. To find the maximum conversion efficiency we use a minimisation algorithm to find the largest value of the conversion efficiency given by Equation $\ref{eq:Cab_sq}$. For this optimisation we are able to vary the atomic and cavity detunings, as well as the microwave and optical coupling rates. For optimising the conversion efficiency we choose $N=1\times10^{16}$ total atoms, all driven by the optical field. This number of atoms corresponds to a crystal 12\,mm in length, driven by a laser with beam width of 0.6\,mm and an erbium ion concentration of 250\,ppm. Atoms driven by the microwave field but not the optical field will lead to an increase in parasitic absorption of the microwave photons, and so minimising the number of atoms driven by only the microwave field will minimise this effect. 

To generate an initial guess for the optimisation algorithm, we first vary the atomic detunings, while keeping the cavity detunings determined by the approximate dressed state detunings and set the coupling rates to the values specified in Table \ref{tab:parameters}. The atomic detunings are varied to find the maximum conversion efficiency, which can be used for our initial guess for the full optimisation process, where the atomic and cavity detunings, and the coupling rates will all be varied. This allows us to find the highest conversion efficiency for given values of other parameters such as laser pump power, intrinsic losses and temperature. 

Increasing the optical laser pump power will increase the conversion efficiency because if there are more optical pump photons then it is more likely that an atom driven into the $\ket2$ state by the input microwaves will be further driven into the $\ket3$ state, which allows for the emission of an upconverted photon. Experimentally, increasing the pump laser power is  limited because higher power will cause heating of the crystal leading to reduced conversion efficiency. The effect of heating can be reduced by pulsing the pump laser with the input microwave photons, rather than have it shining continuously. 

Increasing the quality factor of the optical resonator will increase the conversion efficiency in two ways because both the optical pump and the optical output photons are resonant (with different modes of) the cavity. The higher quality factor will increase the optical pump field strength, which will have a similar effect to increasing the pump power. The intra cavity upconverted field amplitude will also be increased because fewer photons will be lost to intrinsic cavity damping, and so the amplitude of the upconverted output field will be increased, increasing the conversion efficiency. 



\begin{figure}
	\centering
	\includegraphics[width=1\columnwidth]{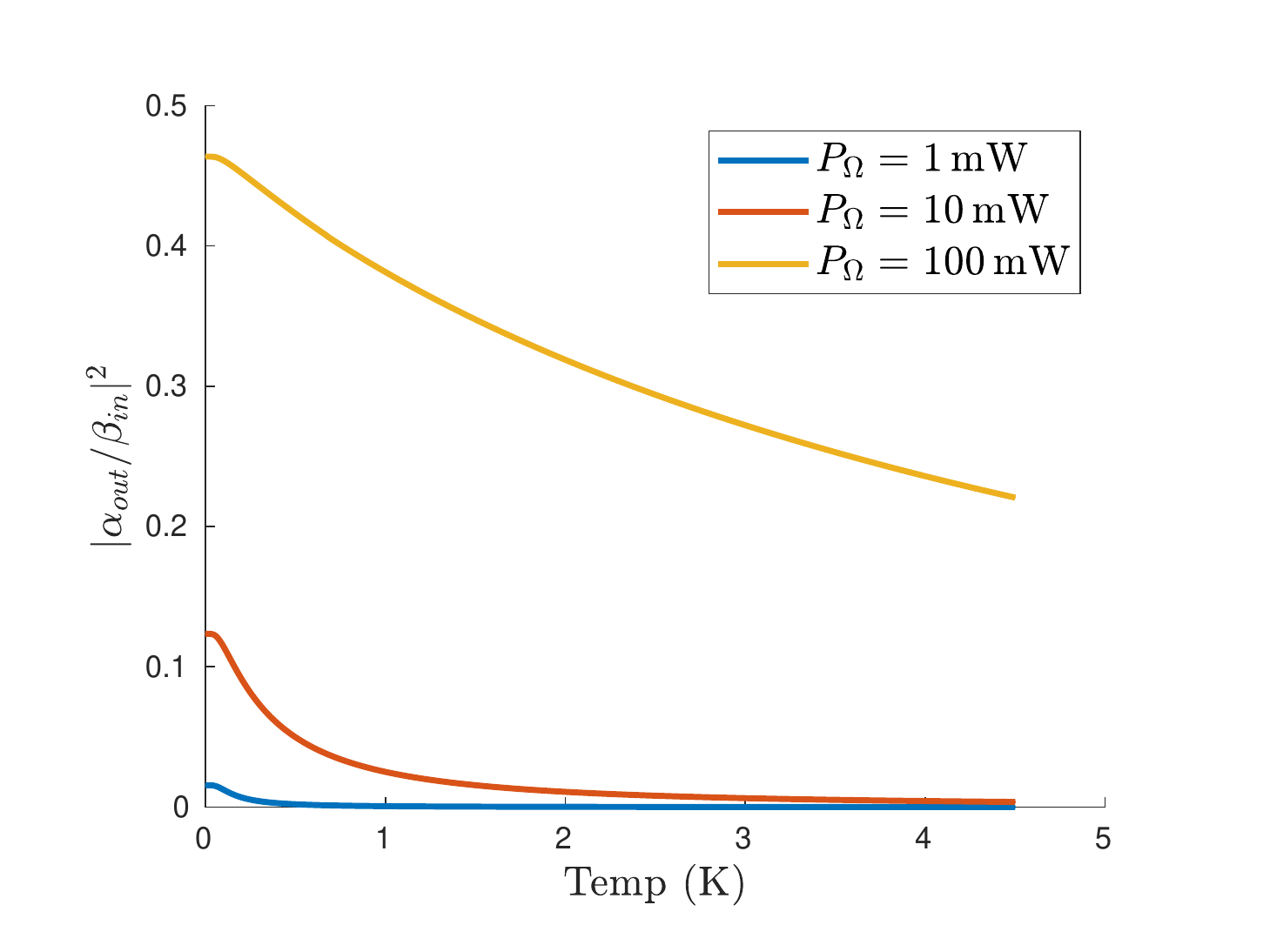}
	\caption{The optimised conversion efficiency as a function of the system temperature, for different values of the pump laser power $P_\Omega$. This simulation was performed using an optical Q factor of $10^8$. }
	\label{fig:effic_vs_temp}
\end{figure}

Figure \ref{fig:effic_vs_temp} shows the effect of temperature on the maximum conversion efficiency for a range of laser pump powers. The conversion efficiency decreases as the temperature increases. For low temperature almost all the ions will be in the $\ket1$ state, and only a small fraction will thermally excited into the $\ket2$ state. The atoms in state $\ket2$ produce a upconversion signal that is out of phase with those in $\ket1$ and to some extent cancels it out. This means that at higher temperatures, where the thermal population of both states is similar, the conversion efficiency is low, especially at low pump powers.

\begin{figure}
	\centering
	\includegraphics[width=1\columnwidth]{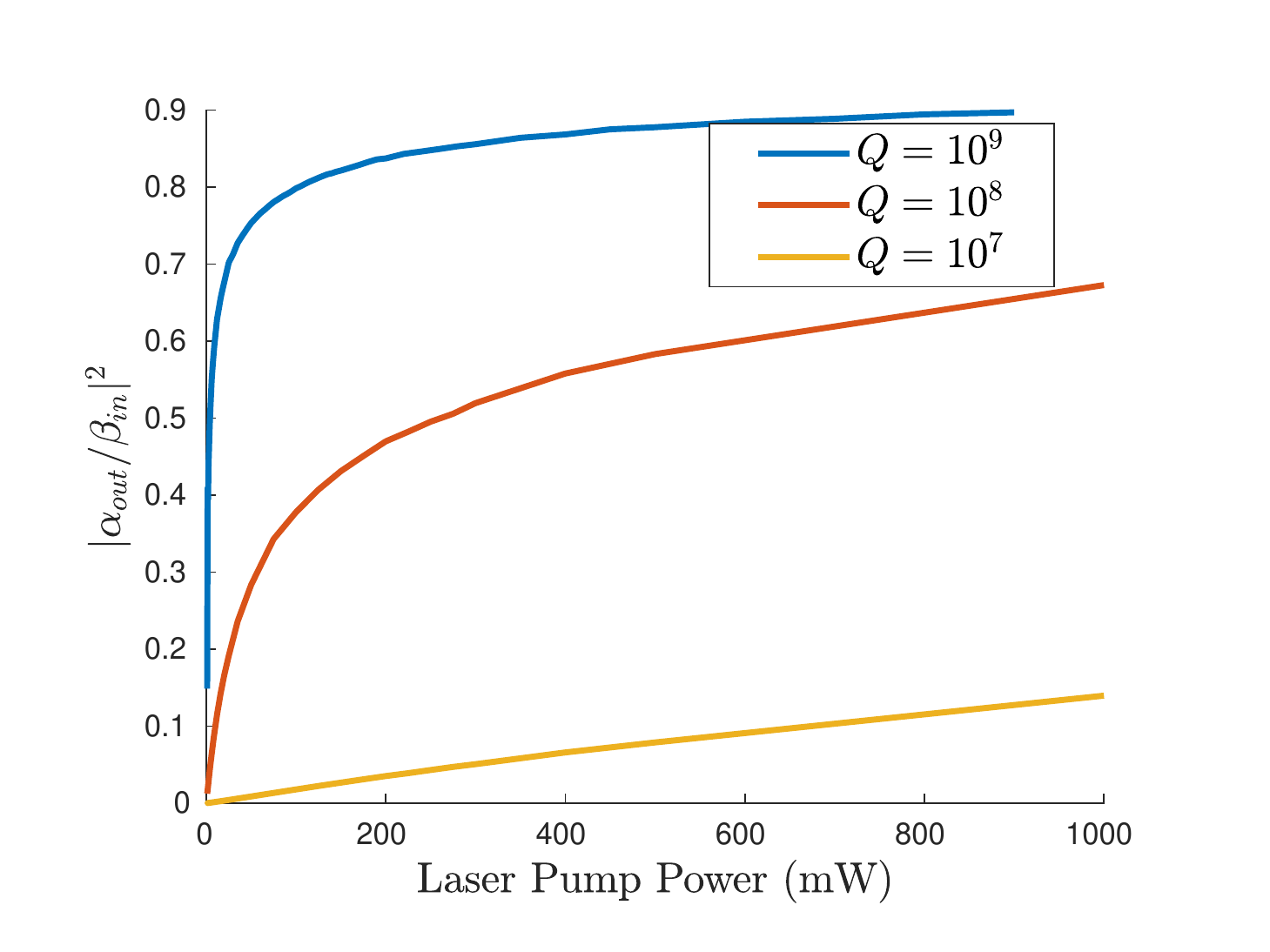}
	\caption{The optimised conversion efficiency as a function of the input laser pump power, so different values of the optical Q factor, at a temperature of 100\,mK.}
	\label{fig:effic_vs_power}
\end{figure}

In Figure \ref{fig:effic_vs_power} the maximum conversion efficiency as been calculated as a function of the optical pump power, for different values of the optical Q factor, showing the conversion efficiency increase as the pump power is increased. For the higher Q factor resonators, initially there are large increases in conversion efficiency as the power is increased, and the rate of increase becomes less at higher pump power. The levelling off occurs when the pump power is high enough such that whenever an ion is excited into the $\ket2$ state by an input microwave photon, it will be further driven into the $\ket3$ state. The conversion efficiency cannot reach 100\% because of the effect of other incoherent loss processes, but by increasing the optical pump power we will decrease the effect of incoherent decay of the $\ket2$ state. For the lower Q resonator ($Q=10^7$) we don't see the same levelling off effect over this range of pump powers because the intra cavity pump field is much smaller, and so increasing the pump power still increases the chance that an ion in state $\ket2$ will further excited into $\ket3$.

Thus with a combination of high Q factor resonators and high optical pump power at low temperatures, attainable using a dilution refrigerator, conversion efficiencies above 80\% should be possible. This is far higher than the largest conversion efficiency found experimentally using this setup which was $1.26\times10^{-5}$. 

The much higher conversion efficiency in these simulations occurs for several reasons. Firstly, the temperature is 100\,mK in the simulations in Figure~\ref{fig:effic_vs_power} which is attainable using a dilution refrigerator, compared with around 4\,K attainable with a cryostat. This lower temperature leads to a much longer lifetime of the microwave transition, so less input microwave photons are lost to incoherent decay. These simulations also assumed a crystal of pure \ce{^{170}Er}, while in the previous experiment the crystal was not isotopically pure, leading to parasitic absorption by \ce{^{167}Er} ions. Finally, in these simulations the detunings and coupling rates were all selected for the highest conversion efficiency.

\section{Conclusion}
We have developed a theoretical description of our rare-earth ion based upconversion device. We began by formulating equations to describe the microwave and optical cavity fields, and from there a description of the interaction between the cavity fields and the inhomogeneously broadened ensemble of ions. Numerical methods were developed to be able to simulate this device. 

From here a simplified model was developed, valid for the quantum information regime where the cavity fields are small. Using realistic experimental parameters for an erbium doped crystal we predict it is possible to reach conversion efficiencies above 80\%. 

\section{Acknowledgements}

The authors would like to thank Ashton Bradley and Gavin King for useful discussions. This work was supported by the US Army Research Office (ARO/LPS) (CQTS) Grant No. W911NF1810011.

%

\bibliographystyle{unsrt}

\end{document}